\documentclass[
    ,final            
  ]
  {aipproc}

\layoutstyle{6x9}

\usepackage{graphicx,amsmath,amsthm,amssymb}

\begin{document}

\title{Single Spin Asymmetries at COMPASS with transverse target polarization}


\author{C. Schill for the COMPASS collaboration}{
  address={Physikalisches Institut der Universit\"at Freiburg\\Hermann-Herder Str. 3\\ D-79104 Freiburg}
}

\begin{abstract} COMPASS is a fixed target experiment at CERN investigating the spin structure of the nucleon and
performing hadron spectroscopy. The transverse spin structure of the nucleon is studied in semi-inclusive
deep-inelastic scattering of $160$~GeV/c muons off a transversely polarized proton or deuteron target. In
$2002$-$2005$, a transversely polarized $^6LiD$, and in $2007$ a transversely polarized $NH_3$ target were used. To get
access to the transversity distribution, different single-spin asymmetries have been measured: the Collins asymmetry,
the hadron-pair asymmetry  and the transverse lambda polarization. In addition, transverse momentum
effects of quarks have been studied by the Sivers effect. New results for the Collins and the Sivers asymmetries on the
proton for  identified pions and kaons and for the two hadron interference asymmetry will be presented. \end{abstract}

\maketitle


\section{Introduction}
\vspace*{-2mm}
The cross-section for semi-inclusive deep-inelastic scattering (SIDIS) in the one-photon exchange approximation
contains eight transverse-momentum dependent distribution functions \cite{Mulders}. Some of these  can be
extracted  measuring the azimuthal distribution of the hadrons in the final state \cite{Collins}. Three
distribution functions  survive upon integration over the transverse momenta: These are the quark momentum
distribution $q(x)$, the helicity distribution $\Delta q(x)$, and the transversity distribution $\Delta_T
q(x)$. The latter is defined as the difference in the number density of quarks with momentum fraction $x$ with
their transverse spin parallel to the the transversely
polarized target and their spin anti-parallel to the target.
\cite{Artru}. \vspace*{-6mm}

\section{The COMPASS experiment}
\vspace*{-2mm}
At COMPASS,  the scattered muon and the produced hadrons are detected in a 50~m long wide-acceptance
forward spectrometer with excellent particle identification capabilities \cite{Experiment}.  A 
transversely polarized   $NH_3$ target consists of three cells aligned along the muon beam axis. The
upstream and downstream cells are polarized in one direction while the middle cell is polarized
oppositely.  The average polarization of the NH$_3$ target is about  $90$\%. The direction of the target
polarization is reversed every few days to reduce the systematic error. The asymmetries are analyzed using at the same time data from
two periods of opposite polarization and  from the different target cells.  Pions and Kaons are
identified in a large scale Ring Imaging Cherenkov Detector (RICH) in a wide momentum range \cite{RICH}.

\section{The Collins asymmetry}

In semi-inclusive deep-inelastic scattering the transversity
distribution $\Delta_Tq(x)$ can be measured in combination with the
chiral odd Collins fragmentation function
$\Delta^0_TD_q^h(x)$. According to Collins, the
fragmentation of a transversely polarized quark into an unpolarized
hadron generates an azimuthal modulation of the hadron distribution 
with respect to the lepton scattering plane \cite{Collins}. The hadron
yield $N(\Phi_{Coll})$ can be written as 
$N(\Phi_{Coll})=N_0\cdot (1+f\cdot P_t\cdot D_{NN}\cdot A_{Coll}\cdot \sin\Phi_{Coll})$, 
where $N_0$ is the average hadron yield, $f$ the fraction of
polarized material in the target, $P_t$ the target polarization,   $A_{Coll}$
the Collins asymmetry, $D_{NN}=(1-y)/(1-y+y^2/2)$ the depolarization factor, and $y$ the fractional 
energy transfer of the muon. The angle $\Phi_{Coll}$ is the so called 
Collins angle. It is defined as  $\Phi_{Coll}=\phi_h-\phi_{s}$, the difference of the
hadron azimuthal angle $\phi_h$ and the quark spin azimuthal angle $\phi_s$ after the scattering, 
both with respect to the lepton scattering plane \cite{Artru}.
 The measured Collins asymmetry $A_{Coll}$ can be factorized into a convolution of the
transversity distribution $\Delta_Tq(x)$ and the
Collins fragmentation function $\Delta_T^0D_q^h(z, p_T)$, summed over all quark
flavors $q$:
\begin{equation}
A_{Coll}=\frac{\sum_q\,  e_q^2\cdot \Delta _Tq(x)\cdot \Delta_T^0D_q^h(z, p_T)}
{\sum_q\, e_q^2 \cdot q(x)\cdot D_q^h(z, p_T)}.
\end{equation}
Here, $e_q$ is the quark charge, $D^h_q(z, p_T)$ the unpolarized fragmentation
function, $z=E_h/(E_\mu-E_{\mu'})$ the fraction of available energy carried by
the hadron and $p_T$ the hadron transverse momentum with respect to the
virtual photon direction. $E_h$, $E_\mu$ and $E_{\mu'}$ are the energies of the
hadron, the muon before and after the scattering, respectively. 

To select DIS events, kinematical cuts on the negative squared four momentum transfer $Q^2>1$~(GeV/c)$^2$, the
hadronic invariant mass $W>5$~GeV/c$^2$ and the fractional energy transfer of the muon $0.1<y<0.9$ are 
applied. The hadron sample on which the single hadron asymmetries are computed consists of all charged
hadrons originating from the reaction vertex with $p_T>0.1$~GeV/c  and $z>0.2$.  The extraction
of the amplitudes is then performed fitting the expression for the transverse polarization dependent part
of the semi-inclusive DIS cross section \cite{Boer} to the measured count rates in the target cells by a
unbinned extended maximum likelihood fit, taking into account the spectrometer acceptance. The results
have been checked by several other methods described in Ref.~\cite{COMPASS}.
\begin{figure}
\includegraphics[width=0.2\textwidth]{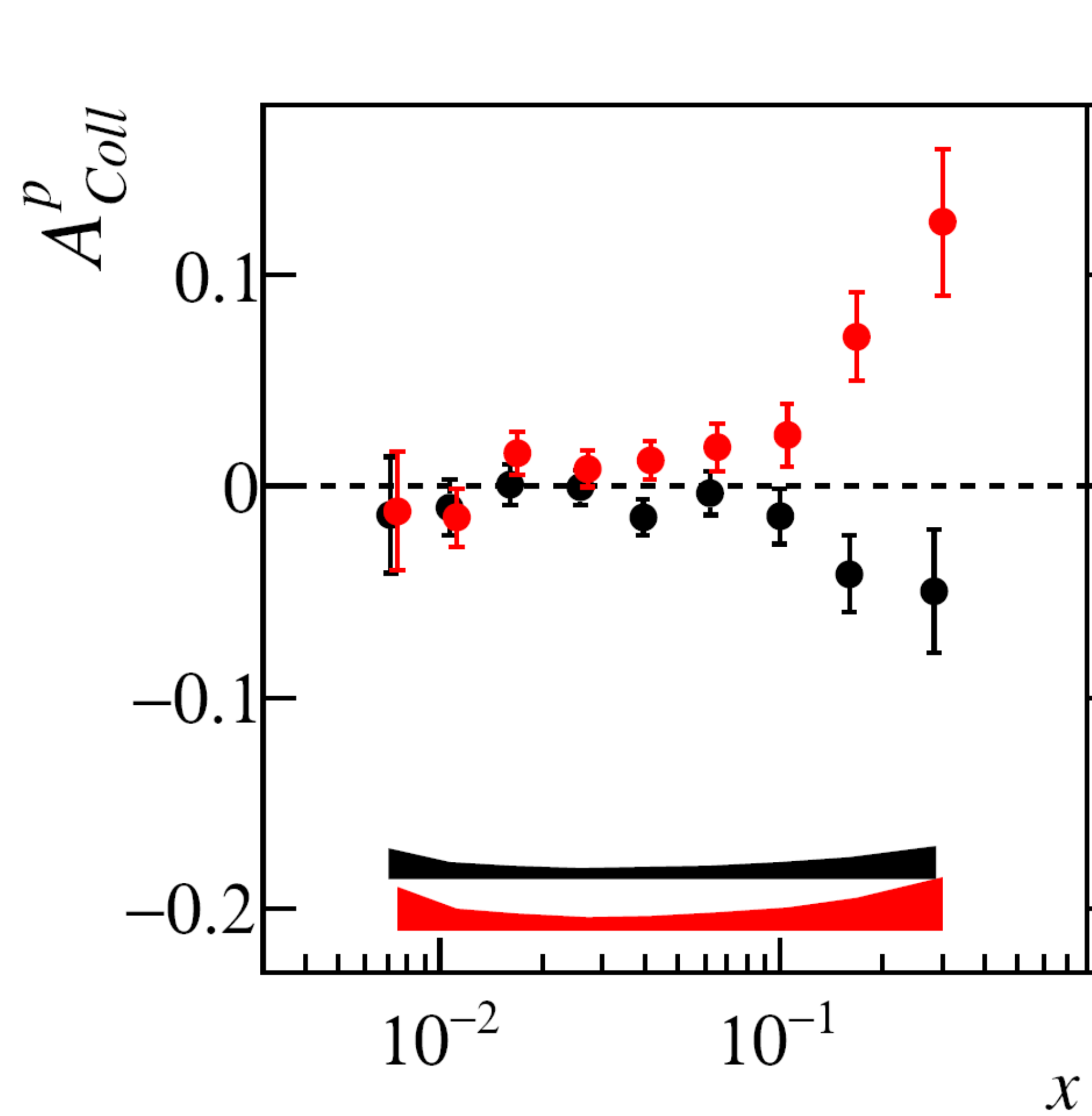}
\includegraphics[width=0.2\textwidth]{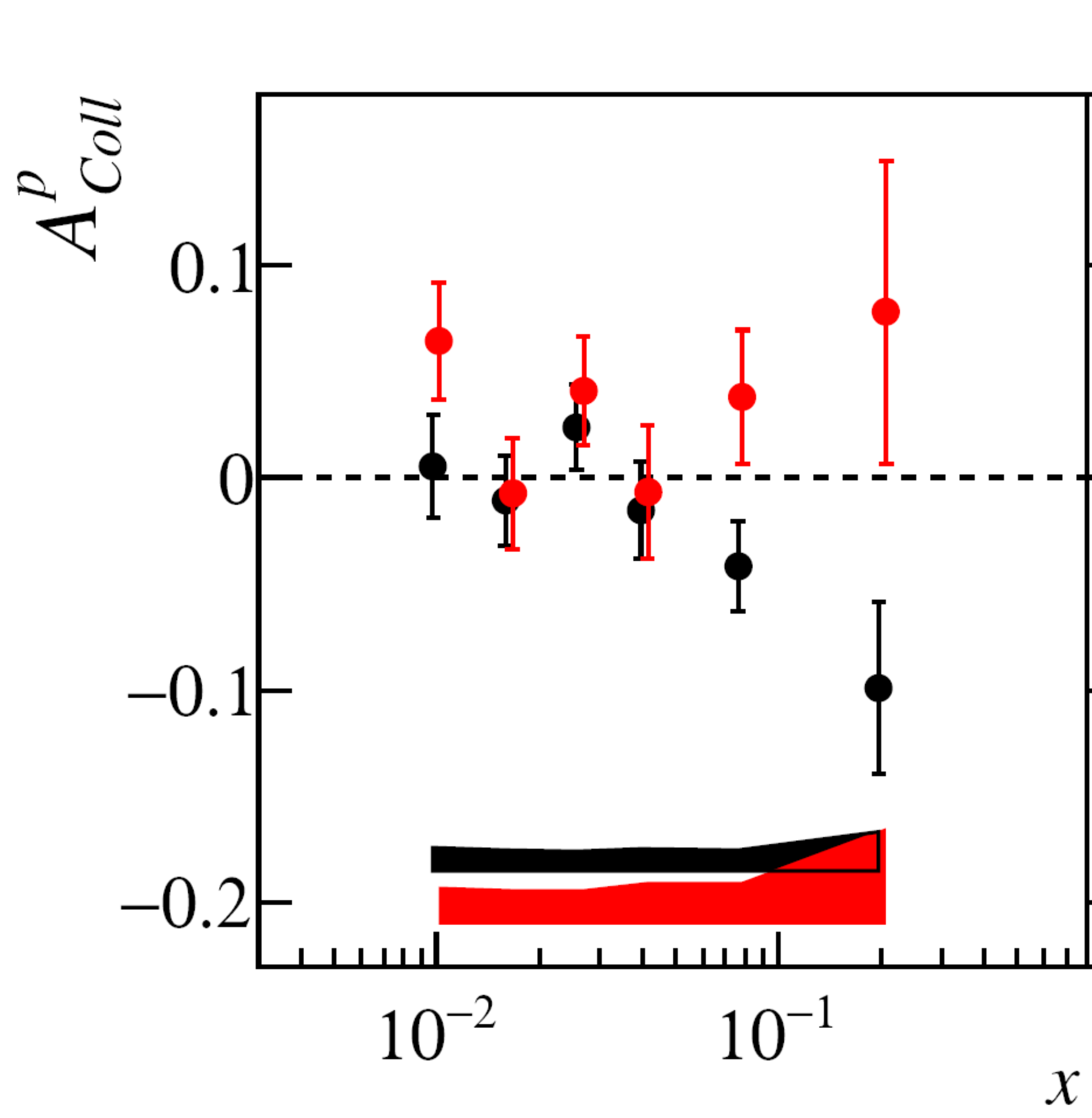}
\includegraphics[width=0.2\textwidth]{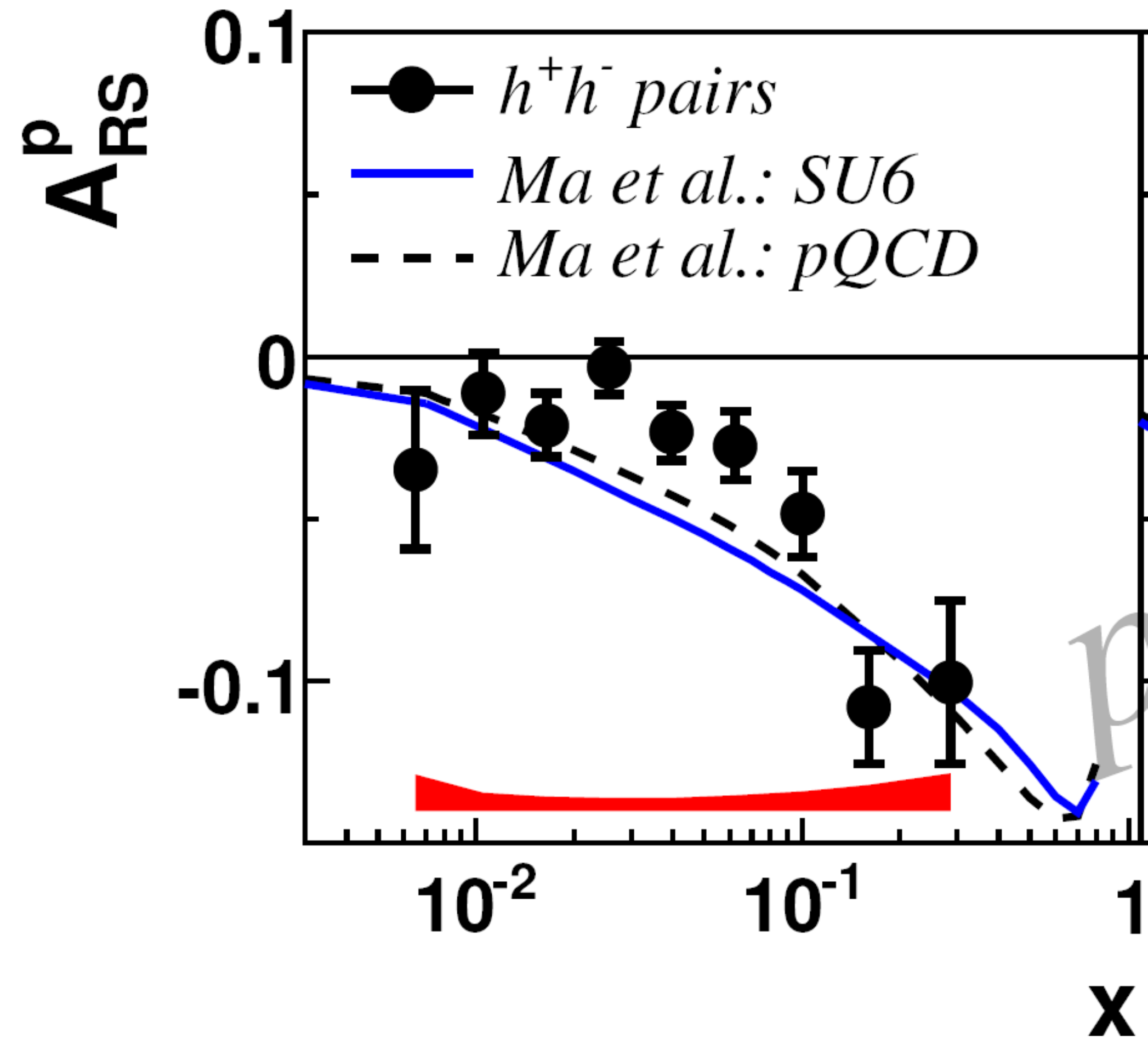}
\includegraphics[width=0.2\textwidth]{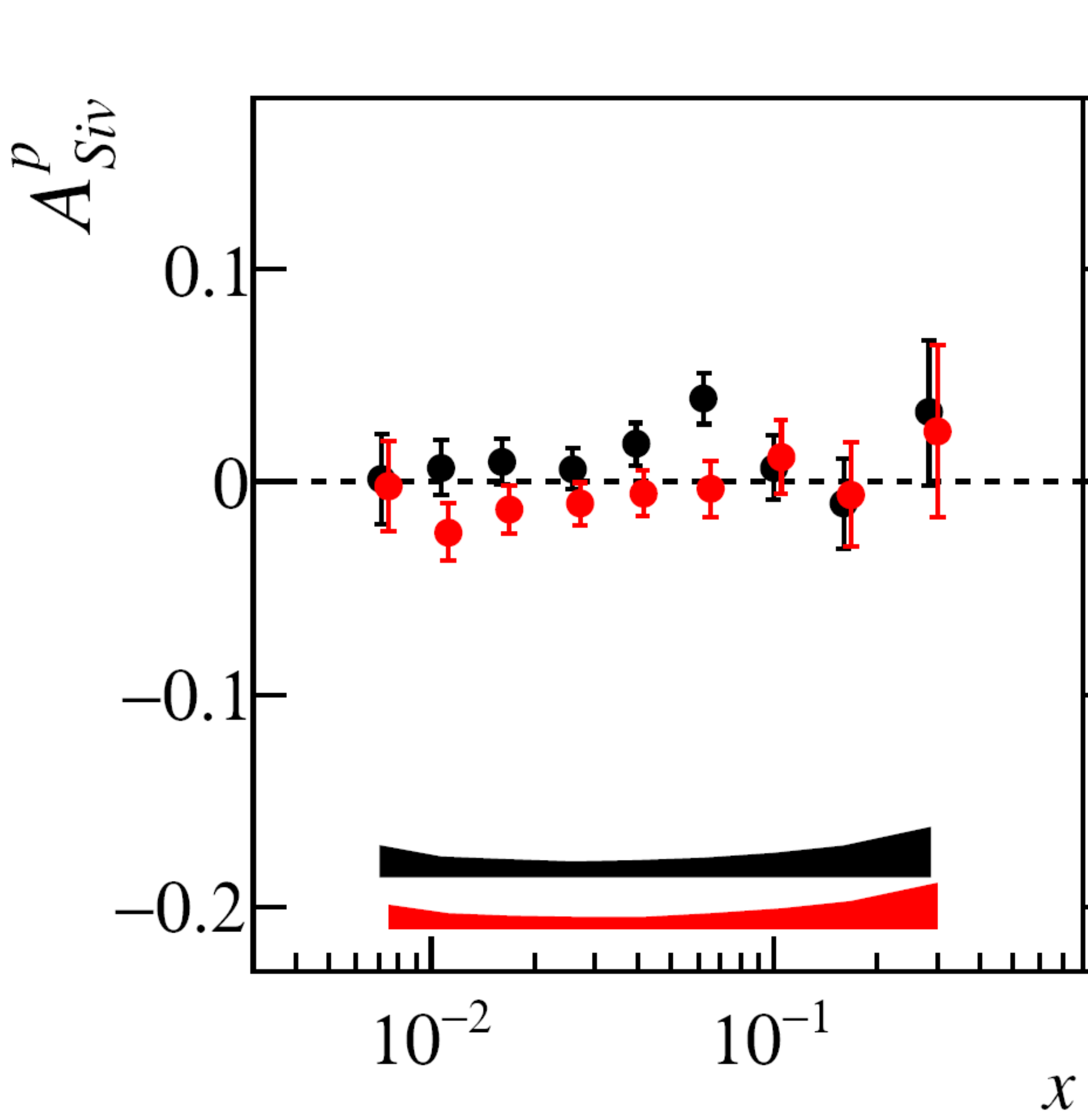} 
\includegraphics[width=0.2\textwidth]{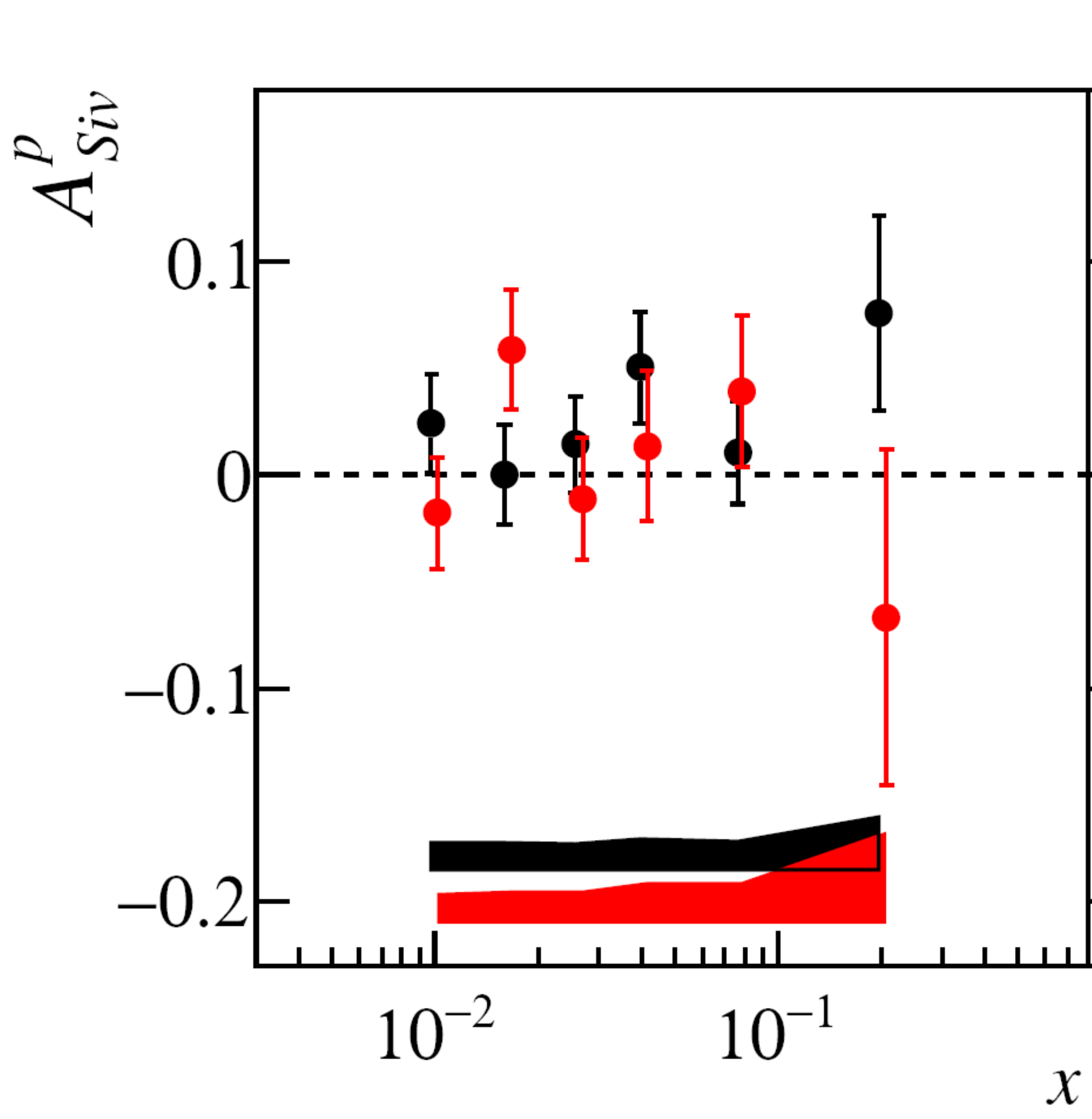}
\caption{Collins asymmetry on the proton for  positive (black) and negative
(red) pions (first panel on the left) and kaons (second panel)  as a function of $x$. 
Two-hadron asymmetry $A_{RS}$ on the proton compared to 
     predictions of \cite{Ma} (third panel).
Sivers Asymmetry for positive (black) and negative (red) pions (fourth panel) and kaons
(right panel).   The bands indicate the
systematic uncertainty of the measurements.}
\label{Collins}
\end{figure}

In left panel of Fig.\ref{Collins} the results for the Collins asymmetry on a $NH_3$ target
are shown as a function of $x$ for positive and negative 
pions. For small $x$ up to $x=0.05$ the measured asymmetry is small and
statistically compatible with zero, while in the last points an asymmetry
different from zero is visible. The asymmetry increases up to about $10$\% with
opposite sign for negative and positive pions. This result confirms the
measurement of a sizable Collins function and transversity distribution. The
asymmetry for positive and negative kaons is shown in the second panel of
Fig.\ref{Collins}.  At larger $x$, the asymmetry is different from zero as well and shows opposite signs
for positive and negative kaons. The data provide important information for global
fits taking into account the Collins fragmentation function from BELLE and the
Collins asymmetries from COMPASS and HERMES to obtain constrains to the
transversity distribution for $u$-, $d$- and $s$-quarks \cite{HERMES, Bacchetta, Efremov,
Aram}.

\section{Two-hadron asymmetry}
\vspace*{-2mm}
The chiral-odd transversity distribution $\Delta_T q(x)$ can also be measured
in combination with the chiral-odd polarized two-hadron interference fragmentation 
function $H^{\sphericalangle}_1 (z,M^2_{inv})$ in SIDIS. $M_{inv}$ is the invariant mass of the
$h^+h^-$ pair. 
The fragmentation of a transversely polarized quark into two unpolarized
hadrons leads to an azimuthal modulation in $\Phi_{RS} = \phi_R - \phi_s$ in the SIDIS cross section. 
Here $\phi_R$ is the azimuthal angle between $\vec R_T$ and the lepton scattering plane and 
$\vec R_T$ is the transverse component of $\vec R$ defined as 
$\vec R = (z_2\cdot \vec p_1 - z_1 \cdot \vec p_2)/(z_1+z_2).$
 $\vec p_1$ and $\vec p_2$ are the momenta in the laboratory frame of $h^+$
and $h^-$ respectively. This definition of $\vec R_T$ is invariant
under boosts along the virtual photon direction. The number of produced oppositely 
charged hadron pairs $N_{h^+h^-}$ can be written as 
$N_{h^+h^-} =N_0 \cdot ( 1 + f \cdot P_t \cdot D_{NN} \cdot A_{RS} \cdot \sin \Phi_{RS} \cdot \sin
\theta)$. Here, $\theta$ is the angle between the momentum vector of $h^+$ in
the center of mass frame of the $h^+h^-$-pair and the momentum vector of
the two hadron system \cite{Bacchetta}. 
 The measured amplitude $A_{RS}$ is proportional to the product of the
transversity distribution and the polarized two-hadron interference fragmentation function 
\begin{equation}
A_{RS} \propto \frac {\sum_q e_q^2 \cdot \Delta_T q(x) \cdot H^{\sphericalangle}_1(z,M^2_{inv})}
 {\sum_q e_q^2 \cdot q(x) \cdot D_q^{2h}(z,M^2_{inv})}.
\end{equation}
$D_q^{2h}(z,M^2_{inv})$ is the unpolarized two-hadron interference fragmentation function.
 For data selection, the hadron pair sample consists of all oppositely charged
hadron pair combinations originating from the reaction vertex. The
hadrons used in the analysis have $z > 0.1$ and $x_F > 0.1$. Both cuts
ensure that the hadron is not produced in the target
fragmentation. To reject exclusively produced $\rho^0$-mesons, a cut on
the sum of the energy fractions of both hadrons was applied
$z_1+z_2<0.9$. Finally, in order to have a good definition of the
azimuthal angle $\phi_R$ a cut on $R_T > 0.07$\,GeV/c was applied. 
The two-hadron asymmetry on the proton as a function of $x$ 
is shown in the third panel of Fig.~\ref{Collins}. A strong asymmetry in
the valence $x$-region is observed, which implies a non-zero transversity
distribution and a non-zero polarized two hadron interference fragmentation
function  $H^{\sphericalangle}_1$. The lines are calculations from Ma {\it et al.}, based on a
SU6 and a pQCD model for transversity \cite{Ma} and the di-hadron fragmentation functions from 
Bacchetta {\it et al.} \cite{Bacchetta}. 
The calculations describe well the magnitude and the $x$-dependence of the measured asymmetry.

\section{The Sivers asymmetry}

Another source of azimuthal asymmetry is related to the Sivers effect. The
Sivers asymmetry rises from a coupling of the intrinsic transverse
momentum $\overrightarrow{k_T}$ of unpolarized quarks with the spin of a
transversely polarized nucleon \cite{Sivers}. The correlation between the transverse nucleon
spin and the transverse quark momentum is described by the Sivers distribution
function  $\Delta_0^Tq(x, \overrightarrow{k_T})$. The Sivers effect results in an
azimuthal modulation of the produced hadron yield: $
N(\Phi_{Siv})=N_0\cdot (1+f\cdot P_t\cdot A_{Siv}\cdot \sin \Phi_{Siv})$. 
The Sivers angle is defined as $\Phi_{Siv}=\phi_h-\phi_S$, where $\phi_S$ is the azimuthal angle of the target spin
vector. The measured Sivers
asymmetry $A_{Siv}$ can be factorized into a product of the Sivers distribution function and the
unpolarized fragmentation function $D_q^h(z)$:
\begin{equation}
A_{Siv}=\frac{\sum_q\, e_q^2\cdot \Delta_0^Tq(x, \overrightarrow{k_T})\cdot  D_q^h(z)}
{\sum_q \,e_q^2\cdot  q(x)\cdot D_q^h(z)}.
\end{equation}
In this case the asymmetry $A_{Siv}$ shows up as the amplitude of a $\sin\Phi_{Siv}$ modulation in the
number of produced hadrons. Since the Collins and Sivers asymmetries are independent azimuthal modulations
of the cross section for semi-inclusive deep-inelastic scattering \cite{Boer},
both asymmetries are determined experimentally in a common fit to the
same dataset, taking into account the acceptance of the spectrometer.

In the fourth panel of Fig.\ref{Collins} the results for the Sivers asymmetry on the proton are shown as a
function of $x$.  The Sivers asymmetry for negative pions   is small and statistically
compatible with zero. For  positive pions the Sivers asymmetry is positive. The Sivers asymmetry for
kaons is shown in the right panel of Fig.\ref{Collins}.

\textbf{Acknowledgments} This work has been supported in part by the German BMBF.


\end{document}